\pdfoutput=1
\documentclass[prx,twocolumn,,floatfix]{revtex4-1}

\usepackage{graphicx}  
\usepackage{dcolumn}   
\usepackage{bm}        
\usepackage{amssymb}   
\usepackage{amsmath}
\usepackage{braket}
\usepackage{mathdots}
\usepackage{bm}
\usepackage[colorlinks=true, linkcolor=blue, urlcolor=blue,
citecolor=blue]{hyperref}

\begin{document}
			
\title{High-harmonic generation in spin and charge current pumping at ferromagnetic or antiferromagnetic resonance in the presence of spin-orbit coupling}		
\author{Jalil Varela Manjarres}
\affiliation{Department of Physics and Astronomy, University of Delaware, Newark, DE 19716, USA}
\author{Branislav K. Nikoli\'{c}}
\email{bnikolic@udel.edu}
\affiliation{Department of Physics and Astronomy, University of Delaware, Newark, DE 19716, USA}
		
\begin{abstract}
One of the cornerstone effects in spintronics is spin pumping by dynamical magnetization that is steadily precessing (around, e.g., the $z$-axis) with frequency $\omega_0$, due to absorption of low-power microwaves of frequency $\omega_0$ under the resonance conditions and in the {\em absence} of any applied bias voltage. The two-decades-old ``standard model'' of this effect, based on the scattering theory of adiabatic quantum pumping, predicts that component $I^{S_z}$ of spin current vector $\big( I^{S_x}(t),I^{S_y}(t),I^{S_z} \big) \propto \omega_0$ is time-independent while $I^{S_x}(t)$ and $I^{S_y}(t)$  oscillate harmonically in time with a {\em single} frequency $\omega_0$; whereas pumped charge current is zero $I \equiv 0$ in the same adiabatic $\propto \omega_0$ limit. Here we employ more general than ``standard model'' approaches, time-dependent nonequilibrium Green's function (NEGF) and Floquet-NEGF, to predict unforeseen  features of spin pumping---precessing localized magnetic moments within ferromagnetic metal (FM) or antiferromagnetic metal (AFM), whose conduction electrons are exposed to spin-orbit coupling  (SOC) of either intrinsic or proximity origin, will pump both spin $I^{S_\alpha}(t)$ and charge $I(t)$ currents. All four of these functions  harmonically  oscillate in time at {\em both even an odd integer multiples} $N\omega_0$ of the driving frequency $\omega_0$. The cutoff order of such high-harmonics increases with SOC strength, reaching  $N_\mathrm{max} \simeq 11$ in the chosen-for-demonstration one-dimensional FM or AFM models. Higher cutoff $N_\mathrm{max} \simeq 25$ can be achieved in realistic two-dimensional (2D) FM models defined on the honeycomb lattice, where we provide prescription on how to realize them using 2D magnets and their heterostructures.
\end{abstract}

\maketitle
  
{\em Introduction}.---The pumping of electronic spin current by dynamical magnetization of a ferromagnetic  metal (FM)  into adjacent normal metal (NM)  was discovered~\cite{Tserkovnyak2005,Ando2014} originally for steadily precessing magnetization of FM due to the absorption of {\em low-power} (\mbox{$\sim$ mW}~\cite{Fan2010}) microwaves in \mbox{$\sim$ GHz} range under the ferromagnetic (F) resonance conditions. Since it occurs in the absence of any bias voltage, it is termed ``pumping'' akin to low-temperature quantum transport  where time-dependent quantum systems emit currents~\cite{Switkes1999,Brouwer1998,Bajpai2019}. Spin pumping has turned out to be a ubiquitous phenomenon in room-temperature spintronic devices, emerging whenever dynamics of localized magnetic moments is initiated while they interact with conduction electrons to drive them out of equilibrium. For example, recent observations include spin pumping into adjacent NM from microwave-driven ferro- and ferri-magnetic insulators~\cite{Yang2018}, antiferromagnetic (AF) insulators~\cite{Vaidya2020,Li2020}, as well as from  dynamical noncollinear magnetic texture (such as domain walls~\cite{Zhang2009,Weindler2014,Petrovic2018,Petrovic2021}, skyrmions~\cite{Abbout2018} and spin waves~\cite{Suresh2020}).   

Despite being quintessentially a quantum transport phenomenon, spin pumping is observed even at room temperature because of its  interfacial nature~\cite{Tserkovnyak2005} where the relevant region around magnetic-material/NM interface is always thinner~\cite{Chen2009} than decoherence lengths for electronic orbital and spin degrees of freedom. Thus, the ``standard model''~\cite{Tserkovnyak2005} of spin pumping is built using the scattering theory~\cite{Brouwer1998} of quantum transport to describe how magnetization of ferromagnets, or both the N\'{e}el vector and nonequilibrium magnetization of  antiferromagnets~\cite{Vaidya2020,Li2020,Cheng2014}, precessing with frequency $\omega_0$ around the easy ($z$-axis) pushes electrons out of equilibrium. The ensuing flowing electronic spins comprise  spin current vector \mbox{$\big(I^{S_x}(t),I^{S_y}(t),I^{S_z}\big)$} whose $I^{S_z}$ component is  time-independent  while $I^{S_x}(t)$ and $I^{S_y}(t)$ oscillate harmonically in time at a {\em single} frequency $\omega_0$~\cite{Tserkovnyak2005}. These DC and AC components of pumped spin current can be converted into DC~\cite{Saitoh2006} and AC~\cite{Wei2014} voltages, respectively, by the inverse spin Hall effect and then measured using standard electrical circuits~\cite{Ando2014}. The magnitude of each component is $\propto \omega_0$, as the signature of general adiabatic quantum  pumping~\cite{Switkes1999,Brouwer1998,Bajpai2019}, as well as $\propto \sin^2 \theta$~\cite{Tserkovnyak2005,Chen2009} where the precession cone angle $\theta$ is controlled by the input microwave power~\cite{Fan2010,Jamali2015}. The spin pumping generates effectively additional  dissipation~\cite{Tserkovnyak2005,Zhang2009,Petrovic2021} for the magnetization dynamics, so that such loss of spin angular momentum can be employed  for indirect~\cite{Tserkovnyak2005,Ando2014,Weindler2014} detection of pumping.

\begin{figure}
	\centering
	\includegraphics[scale=0.42]{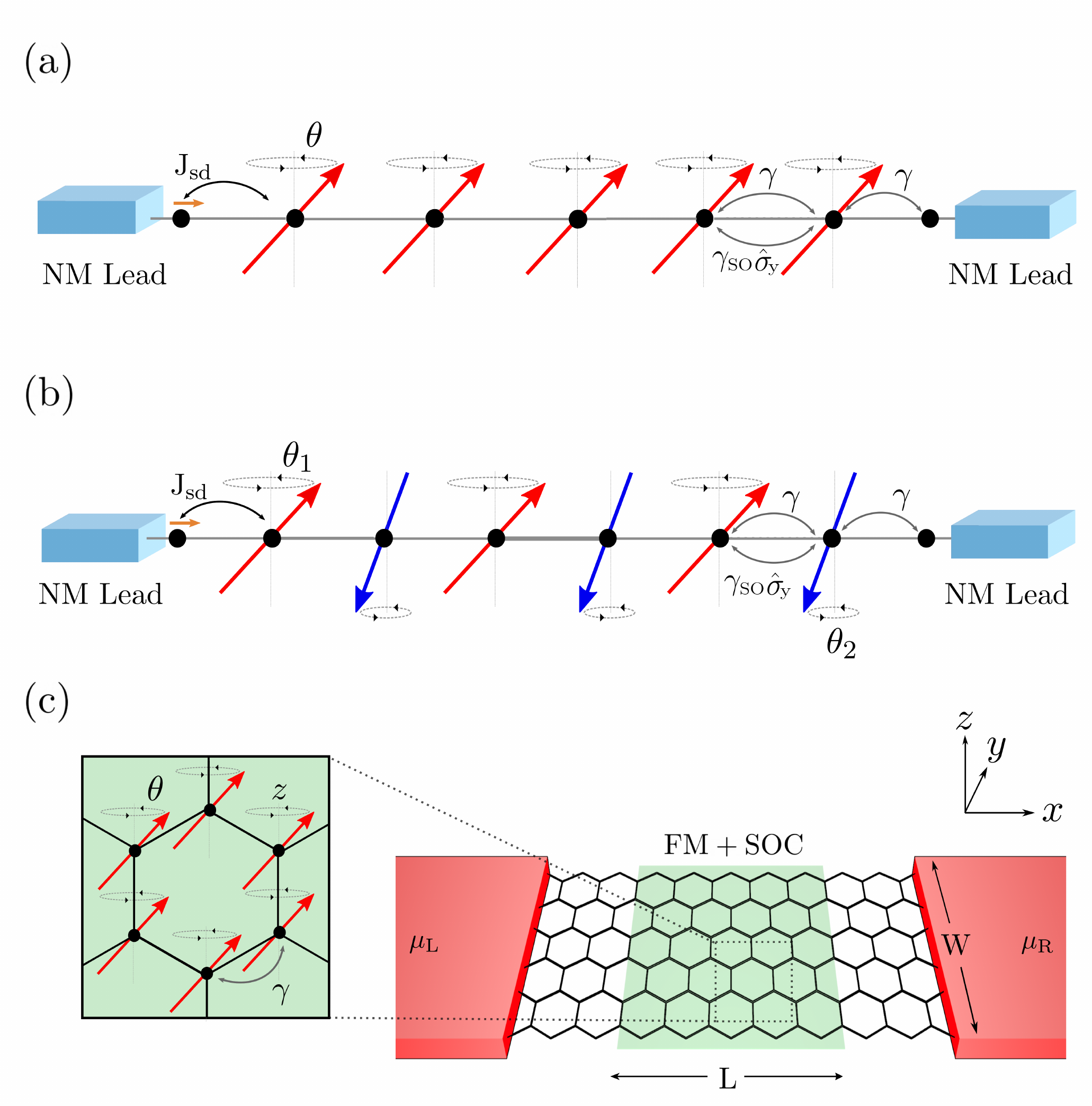}
	\caption{Schematic view of (a),(c) FM- or (b) AFM-based setups whose magnetic moments are steadily precessing with frequency $\omega_0$ around the $z$-axis due to resonant absorption of $\sim $ GHz or $\sim$ THz microwaves, respectively. The setups in (a) and (b) are modeled on 1D TB lattice attached to two semi-infinite NM leads, while setup in (c) in modeled on honeycomb lattice of 2D magnets~\cite{Gibertini2019,Olsen2019,Vanherck2020} and attached to two semi-infinite graphene nanoribbon leads. The precession cone angle  is $\theta$ in the FM case; or $\theta_1 > \theta_2$ for two sublattices in the RH mode of precession~\cite{Vaidya2020,Cheng2014} in the AFM case. In the absence of any bias voltage between macroscopic reservoirs into which NM leads terminate, dynamical magnetic moments interact with conduction electrons via $sd$ exchange coupling $J_{sd}$ to drive them out of equilibrium and pump time-dependent spin $I^{S_\alpha}(t)$ and charge $I(t)$ currents into the NM leads. Electrons hop between the sites with parameter $\gamma$, as well as with an additional [Eqs.~\eqref{eq:hamil} and ~\eqref{eq:hamil2d}] spin-dependent hopping $\gamma_\mathrm{SO}$~\cite{Nikolic2006} describing the Rashba SOC~\cite{Manchon2015} within the FM or AFM central region.}
	\label{fig:fig1}
\end{figure}

Thus, the ``standard model'' of spin pumping {\em excludes} possibility of higher harmonics in periodic time-dependence of 
any of the three components of thus generated spin current. Alternatives to ``standard model'' include Floquet nonequilibrium Green's function  (Floquet-NEGF)~\cite{Mahfouzi2012,Mahfouzi2014,Dolui2020a,Dolui2022,Ahmadi2017} or the Kubo formalism~\cite{Chen2015}, developed in order to include possibly strong spin-orbit coupling (SOC) directly at the F(AF)-material/NM interface where the analytical formula of the ``standard model'' ceases to be applicable~\cite{Tserkovnyak2005,Mahfouzi2012,Dolui2020a,Liu2014}. However, these alternatives have focused on computing time-averaged (i.e.,  DC) component of pumped spin or charge currents, $I^{S_\alpha,\mathrm{DC}}=\frac{1}{\tau} \int_0^\tau\! dt\, I^{S_\alpha}(t)$ or  $I^{\mathrm{DC}}=\frac{1}{\tau} \int_0^\tau\! dt\, I(t)$ where $\tau=2\pi/\omega_0$ is the period, so possible high-harmonics in periodic time-dependence of $I^{S_\alpha}(t)$ ($\alpha \in \{x,y,z\}$) or $I(t)$ are overlooked by them as well. The same focus on DC component of current appears in studies of pumping from nonmagnetic systems~\cite{Arrachea2006}. Extending scattering~\cite{Moskalets2011} or Floquet-NEGF~\cite{Shevtsov2013} formalisms to obtain complete periodic time-dependence, \mbox{$I^{S_\alpha}(t)=I^{S_\alpha}(t+ \tau)$} and \mbox{$I(t)=I(t+\tau)$}, can detect integer high-harmonics, but it would miss possibility of non-integer~\cite{Schmid2021} harmonics (or interharmonics in engineering literature~\cite{Lin2014}).  

\begin{figure}
	\centering
	\includegraphics[scale=0.27]{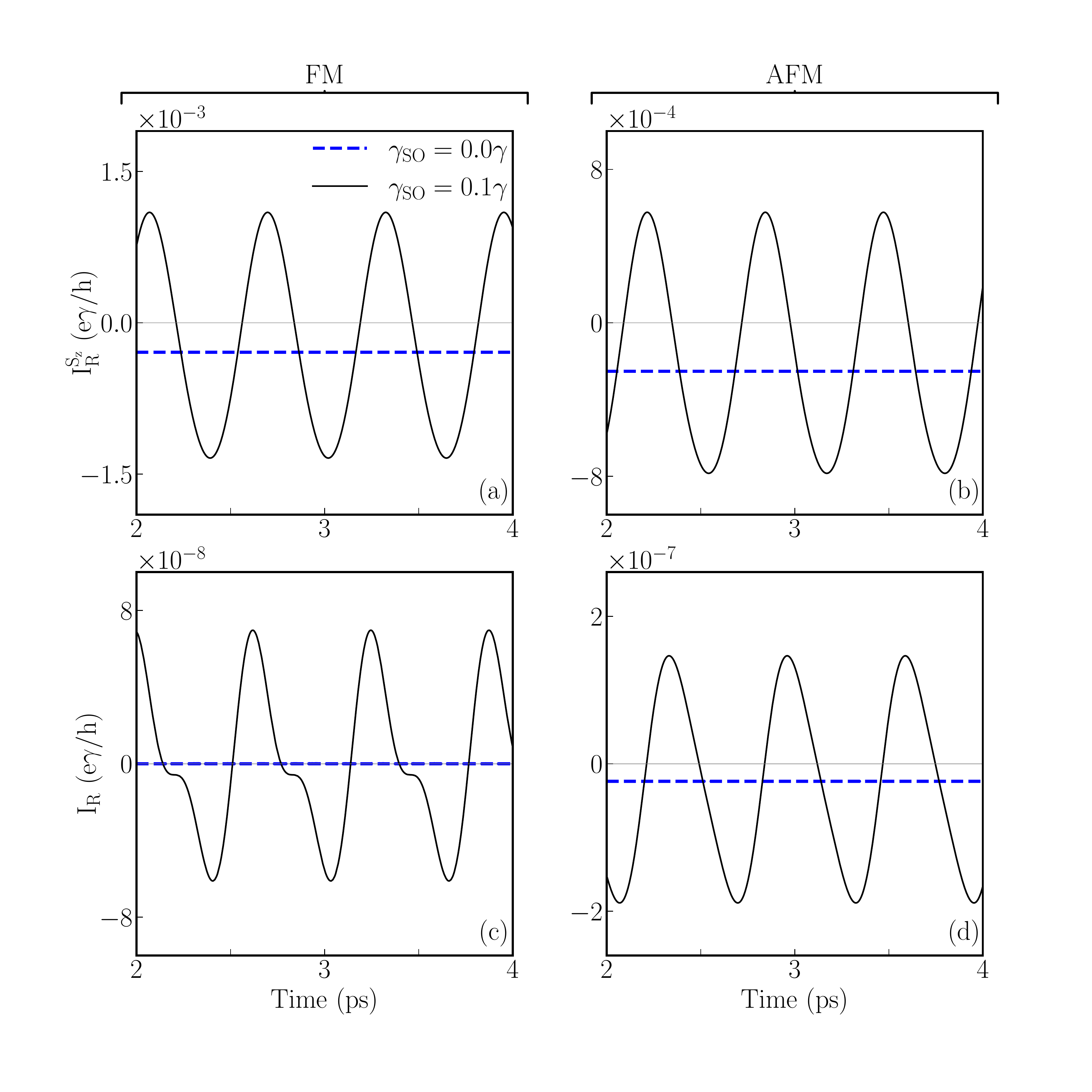}
	\caption{TDNEGF-compute time-dependence of (a),(b) spin $I^{S_z}_R(t)$ and (c),(d) charge $I_R(t)$ currents, after the transient response has died out, pumped into the right NM lead in (a),(c) FM setup  or (b),(d) AFM setup. The Rashba  SOC [Eq.~\eqref{eq:hamil}] is turned off ($\gamma_\mathrm{SO}=0$) or turned on ($\gamma_\mathrm{SO}=0.1 \gamma=J_{sd}$)  within the entire magnetic central region in Fig.~\ref{fig:fig1}.}
	\label{fig:fig2}
\end{figure}

In this Letter, we examine two-terminal one-dimensional (1D) setups illustrated in Fig.~\ref{fig:fig1}(a) and Fig.~\ref{fig:fig1}(b) via time-dependent NEGF (TDNEGF) formalism~\cite{Gaury2014}. In addition, we examine two-terminal two-dimensional (2D) setup in Fig.~\ref{fig:fig1}(c) via time-independent Floquet-NEGF~\cite{Mahfouzi2012,Moskalets2011}. The TDNEGF formalism is more general than standardly used scattering approach to spin pumping~\cite{Tserkovnyak2005,Cheng2014} or Floquet-NEGF formalism as it can yield both transient and (at longer times) time-periodic (or nonperiodic when magnetic moments are not steadily precessing~\cite{Petrovic2018,Petrovic2021}) pumped spin and charge currents  in {\em numerically exact} fashion. However, it is also more expensive computationally and not really necessary if only integer harmonics are confirmed in time-periodic currents in the long time limit. In Fig.~\ref{fig:fig1}(a) FM, and in  Fig.~\ref{fig:fig1}(b) antiferromagnetic metal (AFM), host both magnetic moments and conduction electrons subject to the Rashba type of SOC~\cite{Manchon2015} throughout the whole magnetic region.  These FM or AFM regions are sandwiched between two semi-infinite NM leads terminating into macroscopic reservoirs kept at the same chemical potential $\mu_L=\mu_R=E_F=0$ (so,  lattices in Fig.~\ref{fig:fig1} are half filled by electrons) and temperature \mbox{$T=300$ K}~\cite{Popescu2016}.  In Fig.~\ref{fig:fig1}(c), we assume that 2D FM region---typically defined on the honeycomb lattice and hosting strong SOC~\cite{Gibertini2019,Olsen2019,Vanherck2020} chosen here to be also of the 
Rashba type---is sandwiched between two semi-infinite graphene leads. Starting from equilibrium state described by grand canonical density matrix (see Eq.~(4) in Ref.~\cite{Petrovic2018}), all magnetic moments modeled as classical vectors $\mathbf{M}_i$ of unit length   start to precess uniformly at time $t=0$ around the $z$-axis with frequency $\omega_0$ and precession cone angle $\theta$ in the case of FMs in Figs.~\ref{fig:fig1}(a) and ~\ref{fig:fig1}(c); or with precession cone angles $\theta_1$ and $\theta_2$ for magnetic moments $\mathbf{M}_i$ on sublattice $A$ and $\mathbf{M}_{i+1}$ on sublattice $B$ of AFM.

\begin{figure}
	\centering
	\includegraphics[scale=0.26]{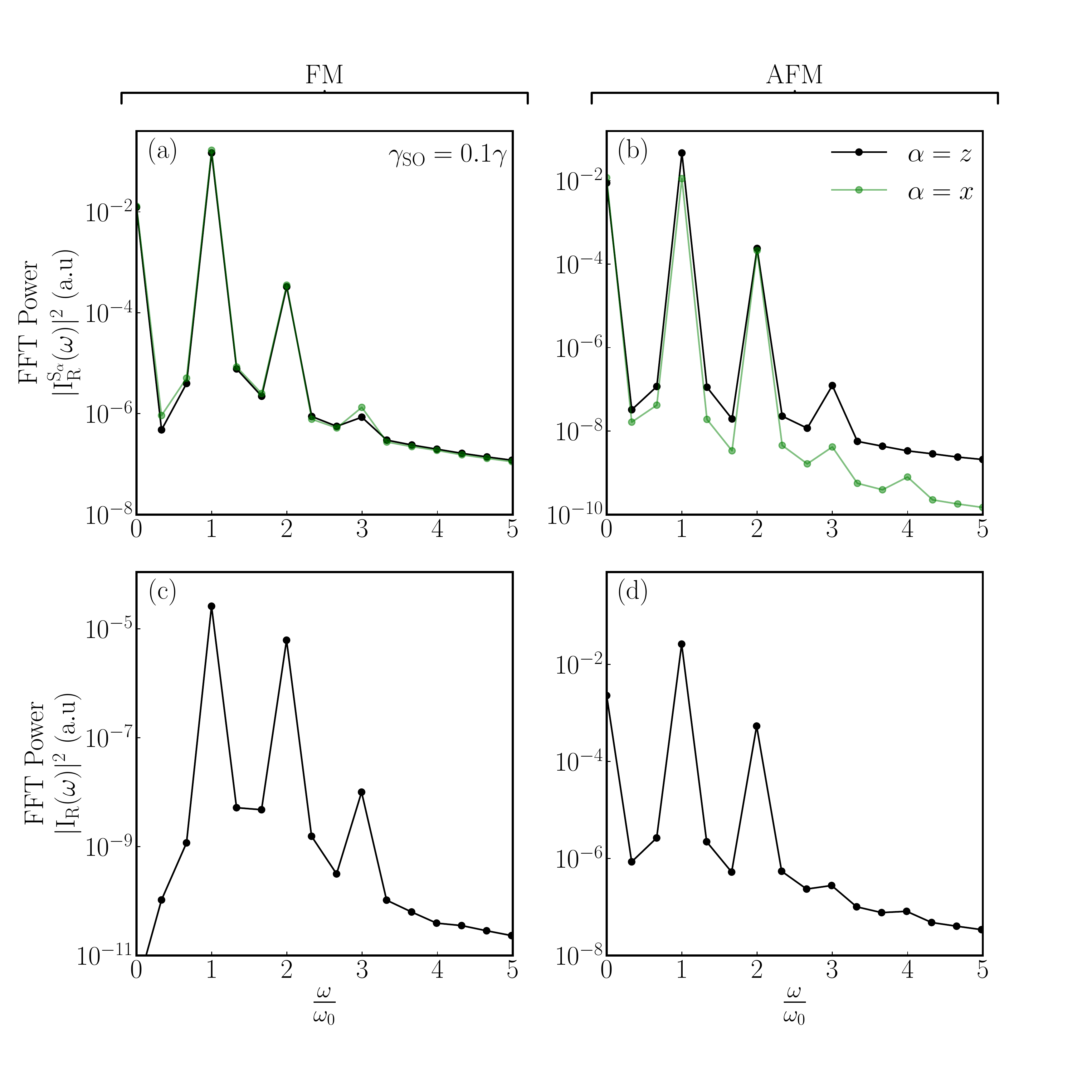}
	\caption{(a)--(d) FFT power spectrum of TDNEGF-computed pumped spin  $|I^{S_z}_R(\omega)|^2$ and charge  $|I_R(\omega)|^2$  currents from   Figs.~\ref{fig:fig2}(a)--(d), respectively. Panels (a) and (b) also show FFT power spectrum of pumped spin current $|I^{S_x}_R(\omega)|^2$.}
	\label{fig:fig3}
\end{figure}

\begin{figure}
	\centering
	\includegraphics[scale=0.26]{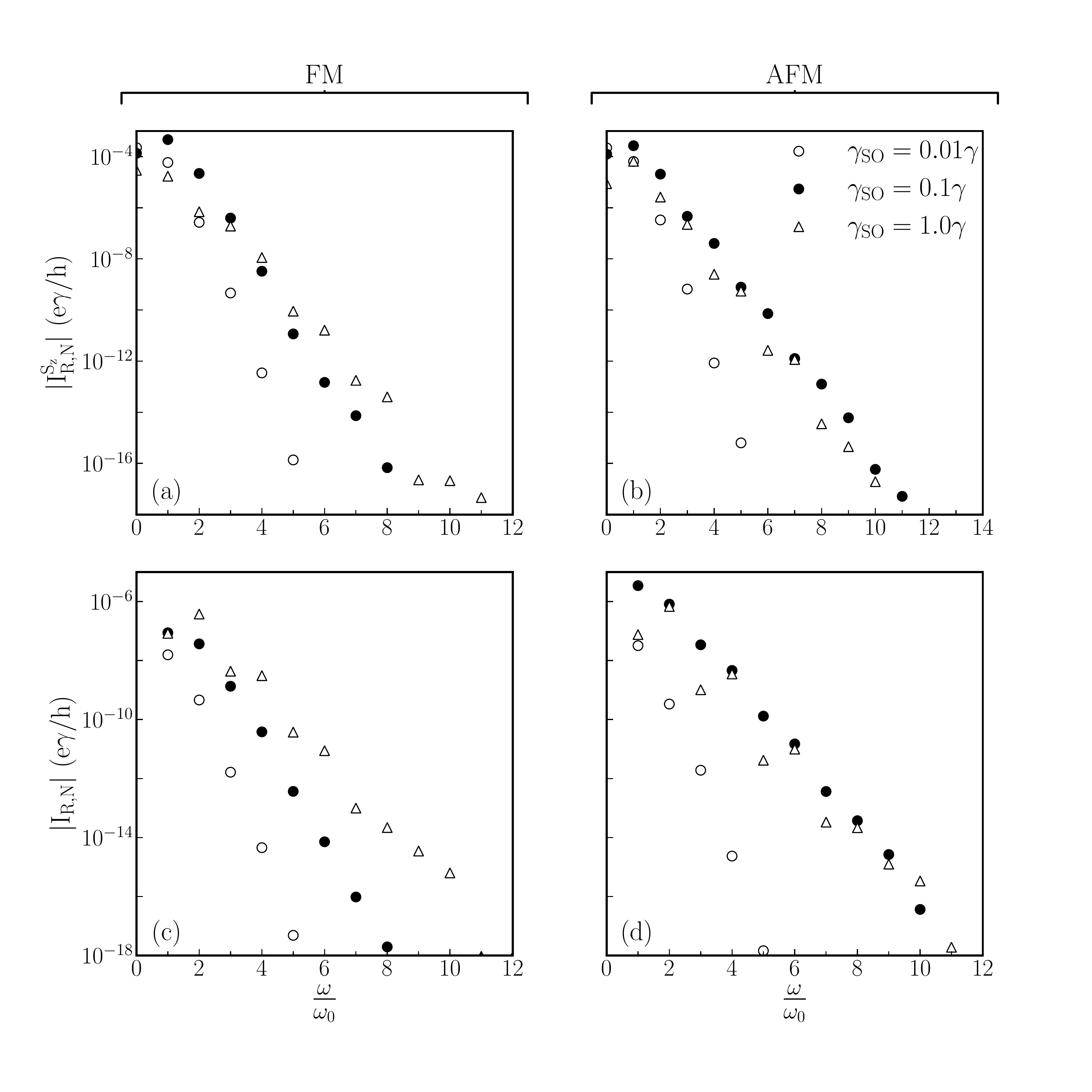}
	\caption{(a)--(d) High harmonics of Floquet-NEGF-computed [Eq.~\eqref{eq:currentfloquet}] pumped spin  $|I^{S_z}_{R,N}|$ and charge  $|I_{R,N}|$  currents for the same 1D FM and AFM systems studied in Figs.~\ref{fig:fig2}(a)--(d), respectively, for three different values of the Rashba SOC [Eq.~\eqref{eq:hamil}] $\gamma_\mathrm{SO}$.}
	\label{fig:fig4}
\end{figure}

\begin{figure}
	\centering
	\includegraphics[scale=0.235]{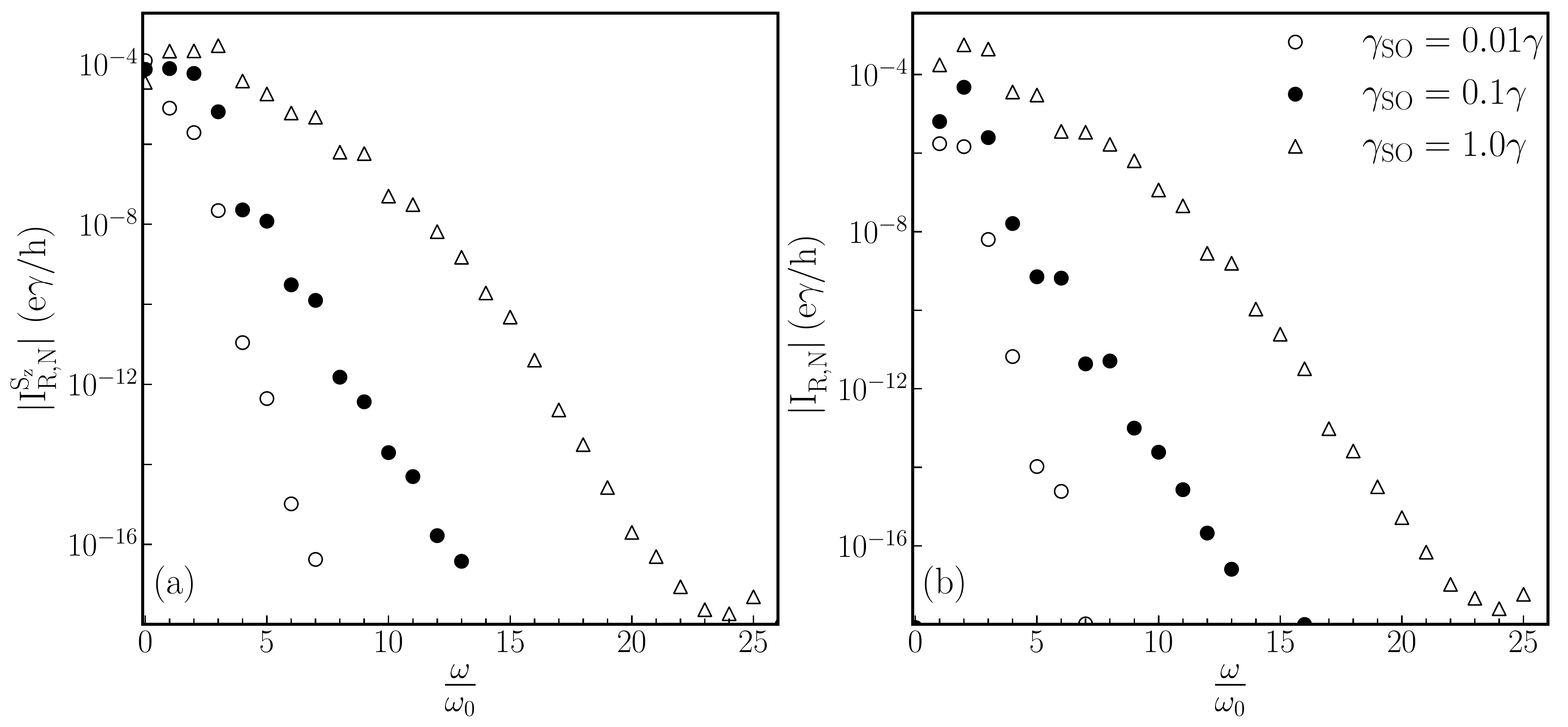}
	\caption{(a)--(d) High harmonics of Floquet-NEGF-computed [Eq.~\eqref{eq:currentfloquet}] pumped spin (a) $|I^{S_z}_{R,N}|$ and (b) charge  $|I_{R,N}|$  currents for 2D FM system on the honeycomb lattice [Fig.~\ref{fig:fig1}(c)] for three different values of the Rashba SOC [Eq.~\eqref{eq:hamil2d}] $\gamma_\mathrm{SO}$.}
	\label{fig:fig5}
\end{figure}

Our principal results in Figs.~\ref{fig:fig2}---\ref{fig:fig5} for pumping at F or AF resonance show that, once the Rashba SOC is turned on, $I^{S_z}(t)$ becomes time-dependent [Figs.~\ref{fig:fig2}(a) and ~\ref{fig:fig2}(b)] and pumping of nonzero charge current $I(t)$ [Figs.~\ref{fig:fig2}(c) and ~\ref{fig:fig2}(d)] ensues as well. Both spin and charge currents exhibit high-harmonics in their fast Fourier transform (FFT) power spectrum [Fig.~\ref{fig:fig3}] at {\em both} even and odd integer multiples $N\omega_0$ of the driving frequency $\omega_0$ whose cutoff $N_\mathrm{max}$ can be controlled by the strength of the SOC [Figs.~\ref{fig:fig4} and ~\ref{fig:fig5}]. In particular, using 2D FM cutoff can reach $N_\mathrm{max} \simeq 25$. Prior to delving into these results, we introduce useful concepts and notation. 


{\em Models and methods}.---The electronic system within FM region in Fig.~\ref{fig:fig1}(a) or AFM region in Fig.~\ref{fig:fig1}(b) is modeled by a 1D tight-binding (TB) Hamiltonian
\begin{eqnarray}\label{eq:hamil}
	\hat{H}_\mathrm{1D}(t) & = & - \gamma \sum_{\langle ij \rangle}  \hat{c}_{i}^{\dagger}  \hat{c}_{j}  - J_{sd}\sum_i\hat{c}_i^\dagger \hat{\bm \sigma} \cdot \mathbf{M}_i(t) \hat{c}_i \nonumber \\ 
	\mbox{} && - i\gamma_\mathrm{SO} \sum_{\langle ij \rangle}  \hat{c}_{i}^{\dagger}  \hat{\sigma}_y \hat{c}_{j},
\end{eqnarray}
where  \mbox{$\hat{c}_i^\dagger=(\hat{c}_{i\uparrow}^\dagger \  \ \hat{c}_{i\downarrow}^\dagger)$} is a row vector containing operators $\hat{c}_{i\sigma}^\dagger$ which create an electron with spin $\sigma=\uparrow,\downarrow$ at site $i$; $\hat{c}_i$ is a column vector containing the corresponding annihilation operators; $\gamma$ is the hopping between the nearest-neighbor (NN) sites (signified by $\langle \ldots \rangle$), also setting the unit of energy; $\gamma_\mathrm{SO}$ is an additional spin-dependent hopping~\cite{Nikolic2006} due to the Rashba SOC~\cite{Manchon2015}; and the conduction electron spin, described by the vector of the Pauli matrices \mbox{$\hat{\bm  \sigma} = (\hat{\sigma}_x,\hat{\sigma}_y,\hat{\sigma}_z)$}, interacts with the classical magnetic moments  $\mathbf{M}_i(t)$ via $sd$ exchange interaction of strength \mbox{$J_{sd} = 0.1\gamma$}~\cite{Cooper1967}. We use $N_\mathrm{sites}=9$ ($N_\mathrm{sites}=10$) sites in FM (AFM) central region, ensuring maximum outflowing spin current (which in 1D oscillates as a function of $N_\mathrm{sites}$~\cite{Petrovic2018}). The left ($L$) and the right ($R$) NM leads, sandwiching FM or AFM region in Fig.~\ref{fig:fig1}, are semi-infinite 1D TB chains described by the first term alone in Eq.~\eqref{eq:hamil}. The Fermi energy of the macroscopic reservoirs into which NM leads terminate is $E_F=0$.

Similarly, the electronic system within FM region in Fig.~\ref{fig:fig1}(c) is modeled by a 2D TB Hamiltonian defined on the honeycomb lattice 
\begin{align}\label{eq:hamil2d}
	\hat H_\mathrm{2D}(t) =&-\gamma \sum_{\langle i j\rangle} \hat c_{i }^{\dagger} \hat c_{j }-J_{sd} \sum_{i}\hat c_{i }^{\dagger}\hat{\bm  \sigma} \cdot  \mathbf{M}(t) \hat c_{i } \nonumber \\ 
	&+i \gamma_{\mathrm{SO}} \sum_{\langle i j\rangle}\hat c_{i }^{\dagger}\left( \hat{\bm \sigma} \times \mathbf{d}_{i j}\right) \cdot \mathbf{e}_z \hat c_{j},
\end{align}
where $\mathbf{d}_{i j}$ is is the vector connecting NN sites $i$ and $j$, $\gamma_{\mathrm{SO}}$ is the strength of the Rashba SOC and  $\mathbf{e}_z$ is the unit vector along the $z$-axis. The L and the R leads Fig.~\ref{fig:fig1}(c) are semi-infinite graphene nanoribbons (GNRs) with zigzag edges described by the first term alone in Eq.~\eqref{eq:hamil2d}. In Eq.~\eqref{eq:hamil2d}, we use the same hopping $\gamma$ in the FM region and in GNR leads, while \mbox{$J_{sd} = 0.1\gamma$}  in the FM region only. The size of the FM region is $4.5 \sqrt{a} \times 8 a$, with $a$ being the distance between NN sites.  The Fermi energy of the macroscopic reservoirs into which GNR leads terminate is $E_F=0$.

In the FM cases [Figs.~\ref{fig:fig1}(a) and ~\ref{fig:fig1}(c)], all magnetic moments precess uniformly with the same frequency $\omega_0$ and the cone angle $\theta=20^\circ$ (which is near the maximum that can be achieved in practice without introducing nonlinearities~\cite{Jamali2015,Fan2010}). This means that $\mathbf{M}_i(t)=\big(\sin \theta \cos(\omega_0 t),\sin \theta \sin(\omega_0 t), \cos \theta \big)$ is plugged into Eq.~\eqref{eq:hamil}. On the other hand, at AF resonance two precession modes of sublattice magnetic moments are possible, with left-handed (LH) and right-handed (RH) chiralities~\cite{Cheng2014}, where both $\mathbf{M}_i^A(t)$ and $\mathbf{M}_i^B(t)$ undergo a clockwise or counterclockwise precession with $\pi$ phase difference, respectively. Thus, in the case of the AFM central region  [Fig.~\ref{fig:fig1}(b)] we use the RH mode---$\mathbf{M}_i(t)=\big(\sin \theta \cos(\omega_0 t),\sin \theta \sin(\omega_0 t), \cos \theta \big)$and $\mathbf{M}_{i+1}(t)=\big(\sin \theta \cos(\omega_0 t + \pi),\sin \theta \sin(\omega_0 t + \pi), \cos \theta \big)$---with $\theta_1=20^\circ$ and the ratio $\theta_1/\theta_2 =1.29$ fixed to correspond to RH mode of MnF$_2$ employed in recent experiments~\cite{Vaidya2020}. We use the same driving frequency \mbox{$\hbar \omega_0=0.01\gamma$} for both FM and AFM 1D cases, which is realistic for the latter but too large for the former. This reduces computational expense of TDNEGF calculations, while not affecting the result since the magnitude of all pumped currents scale linearly with $\omega_0$ (so, the results for the FM case are easily rescaled to realistic frequencies). For 2D FM case [Fig.~\ref{fig:fig1}(c)] calculations via Floquet-NEGF formalism, choice 
of frequency is not important computationally, as long as $\hbar \omega \ll E_F$  so that expression for harmonics of pumped current can be simplified to Eq.~\eqref{eq:currentfloquet} not requiring integration~\cite{Moskalets2011} over energy.

The fundamental quantity of quantum statistical mechanics is the density matrix. The time-dependent one-particle nonequilibrium density matrix  can be expressed~\cite{Gaury2014},  ${\bm \rho^{\rm neq}}(t) = \hbar \mathbf{G}^<(t,t)/i$, in terms of the lesser Green's function of TDNEGF formalism  defined by \mbox{$G^{<,\sigma\sigma'}_{ii'}(t,t')=\frac{i}{\hbar} \langle \hat{c}^\dagger_{i'\sigma'}(t') \hat{c}_{i\sigma}(t)\rangle_\mathrm{nes}$} where  
$\langle \ldots \rangle_\mathrm{nes}$ is the nonequilibrium statistical average~\cite{Stefanucci2013}. We solve a matrix integro-differential equation~\cite{Popescu2016,Petrovic2018}  
\begin{equation}\label{eq:rhoneq}
	i\hbar\frac{d {\bm \rho}^{\rm neq}}{dt} = [\mathbf{H}(t),{\bm \rho}^{\rm neq}] + i \sum_{ p= L,R} [{\bm \Pi}_p(t) + {\bm \Pi}_p^{\dagger}(t)],
\end{equation}
for the time evolution of ${\bm \rho}^{\rm neq}(t)$, where $\mathbf{H}(t)$ is the matrix representation of Hamiltonian in Eq.~\eqref{eq:hamil}. Equation~\eqref{eq:rhoneq} is an {\em exact} quantum master equation for the reduced density matrix of the central FM or AFM region viewed as an open finite-size quantum system attached to macroscopic Fermi liquid reservoirs via semi-infinite NM leads.   The ${\bm \Pi}_p(t)$ matrices
\begin{equation}\label{eq:current}
	{\bm \Pi}_p(t) = \int_{t_0}^t \!\! dt_2\, [\mathbf{G}^>(t,t_2){\bm \Sigma}_p^<(t_2,t) 
	- \mathbf{G}^<(t,t_2){\bm \Sigma}_p^>(t_2,t) ],
\end{equation} 
are expressed in terms of the lesser and greater Green's functions~\cite{Stefanucci2013} and the corresponding self-energies ${\bm \Sigma}_p^{>,<}(t,t')$~\cite{Popescu2016}. They yield  directly time-dependent charge current, \mbox{$I_p(t) = \frac{e}{\hbar} \mathrm{Tr}\, [{\bm \Pi}_p(t)]$}, and spin current, \mbox{$I_p^{S_{\alpha}}(t) = \frac{e}{\hbar} \mathrm{Tr}\, [\hat{\sigma}_{\alpha}{\bm \Pi}_p(t)]$}, pumped into the lead $p = L,R$. We use the same units for charge and spin currents, defined as \mbox{$I_p = I_p^{\uparrow} + I_p^{\downarrow}$} and \mbox{$I_p^{S_{\alpha}} = I_p^{\uparrow} - I_p^{\downarrow}$}, in terms of spin-resolved charge currents $I_p^{\sigma}$. In our convention, {\em positive} current in NM lead $p$ means charge or spin is flowing {\em out} of that lead. 

Although we use 1D TB chain to model FM and AFM setups in Fig.~\ref{fig:fig1}, these setups can  be easily converted into three-dimensional (3D) realistic junctions with macroscopic cross section by assuming that chain is disorder-free and periodically repeated in the $y$- and $z$-directions. This means that our TDNEGF calculations would have to be repeated at each $(k_y,k_z)$ point~\cite{Chen2009}. Nevertheless, studying simpler 1D models can capture essential 
features of pumping from realistic 3D systems (e.g., compare Fig.~3 for 3D junction with realistic atomistic structure to Fig.~4 for 1D junction described by simplistic TB model in Ref.~\cite{Dolui2022}).

In addition to TDNEGF calculations, we also employ Floquet-NEGF approach, operating with time-independent quantities [Eqs.~\eqref{eq:doubleft}--\eqref{eq:fisherlee}], which is far less computationally demanding  and also it can be used to validate [Fig.~\ref{fig:fig4}] TDNEGF calculations once one confirms [Fig.~\ref{fig:fig3}] that only integer harmonics are present in the pumped currents. The Floquet theorem is usually discussed~\cite{Shirley1965,Sambe1973,Eckardt2015} as specifying the form of the solution  of the time-dependent Schr\"{o}dinger equation, \mbox{$i\hbar\partial\ket{\psi(t)}/\partial t=\hat H(t)\ket{\psi(t)}$} for Hamiltonian periodic in time \mbox{$\hat H(t+\tau)=\hat H(t)$}. That is,  
arbitrary solution $\ket{\psi(t)} = \sum_\eta c_\eta e^{-i \varepsilon_\eta (t-t_0)} \ket{u_\eta(t)}$ can be expanded in terms of Floquet states $\ket{\phi_\eta(t)}=e^{-i \varepsilon_\eta t/\hbar} \ket{u_\eta(t)}$ with periodic $\ket{u_\eta(t+\tau)}=\ket{u_\eta(t)}$, quasienergy $\varepsilon_\eta \in \mathbb{R}$ and $c_\eta = \braket{u_\eta(t_0)| \psi(t_0)}$. The same Floquet theorem can be re-stated for double Fourier transformed (i.e., from $t,t'$ to $E,E'$ variables) NEGFs
\begin{equation}\label{eq:doubleft}
	{\bf G}^{r,<}(t,t')=\int\limits_{-\infty}^{+\infty} \frac{dE}{2\pi} \! \int\limits_{-\infty}^{+\infty}  \frac{dE'}{2\pi} \, e^{-i E t/\hbar + i E' t'/\hbar} {\bf G}^{r,<}(E,E'),
\end{equation}
as the requirement that energies $E,E'$ are not independent (as is would be the case in general situation where Hamiltonian is not time-periodic) but instead satisfy
\begin{equation}\label{eq:floquetgf}
	{\bf G}^{r,<}(E,E')={\bf G}^{r,<}(E,E+n \hbar \omega_0)={\bf G}^{r,<}_n(E),
\end{equation}
The coupling of energies $E$ and $E+ n\hbar\omega_0$ ($n$ is integer) indicate  ``multiphoton'' exchange processes. In the absence of many-body (electron-electron or electron-boson)  interactions, currents can be expressed using solely~\cite{Mahfouzi2012,Bajpai2020a} the Floquet-retarded-GF $\check{\bold{G}}^r(E)$
\begin{equation}\label{eq:floquet_GF}
	[E + \check{\bold{\Omega}} - \check{\bold{H}}_\mathrm{F} - \check{\bold{\Sigma}}^r(E) ]\check{\bold{G}}^r(E) = \check{\bold{1}},
\end{equation}
which is composed of ${\bf G}^r_n(E)$ submatrices along the diagonal. Here 
\begin{equation}\label{eq:floquet_ham}
	\check{\bold{H}}_\mathrm{F} = \begin{pmatrix}
		\ddots & \vdots & \vdots & \vdots & \iddots\\
		\cdots & \bold{H}_0 & \bold{H}_1 & \bold{H}_2 & \cdots \\
		\cdots & \bold{H}_{-1} & \bold{H}_0 & \bold{H}_1 & \cdots \\
		\cdots & \bold{H}_{-2} & \bold{H}_{-1}& \bold{H}_0 & \cdots \\
		\iddots & \vdots & \vdots & \vdots & \ddots
	\end{pmatrix}.
\end{equation}
is time-independent but infinite matrix representation of the so-called Floquet Hamiltonian~\cite{Shirley1965,Sambe1973,Eckardt2015} whose finite submatrices $\bold{H}_n$ of size $N_\mathrm{sites} \times N_\mathrm{sites}$ are matrix representations (in the basis of orbitals $|i\rangle$) of the coefficients $\hat{H}_n$ in the Fourier expansion $\hat{H}(t) = \sum_{n=-\infty}^{\infty}  e^{-i n \omega_0 t} \hat{H}_n$. In addition, in Eq.~\eqref{eq:floquet_GF} we use the following notation
\begin{equation}\label{eq:omega_mtx}
	\check{\bold{\Omega}} = \begin{pmatrix}
		\ddots & \vdots & \vdots & \vdots & \iddots\\
		\cdots & -\hbar\omega_0 \bold{1} & 0 & 0 & \cdots \\
		\cdots & 0 & 0 & 0 & \cdots \\
		\cdots & 0 & 0 & \hbar\omega_0 \bold{1}& \cdots \\
		\iddots & \vdots & \vdots & \vdots & \ddots
	\end{pmatrix},
\end{equation}
and $\check{\bold{\Sigma}}^r(E)$  is the retarded Floquet self-energy matrix
\begin{equation}\label{eq:floquet_self}
	\check{\bold{\Sigma}}^r(E) = \begin{pmatrix}
		\ddots & \vdots & \vdots & \vdots & \iddots\\
		\cdots & \bold{\Sigma}^r(E-\hbar\omega_0) & 0 & 0 & \cdots \\
		\cdots & 0 & \bold{\Sigma}^r(E)  & 0 & \cdots \\
		\cdots & 0 & 0 & \bold{\Sigma}^r(E+\hbar\omega_0)  & \cdots \\
		\iddots & \vdots & \vdots & \vdots & \ddots
	\end{pmatrix},
\end{equation}
composed of the usual self-energies of the  leads~\cite{Velev2004}, $\bold{\Sigma}^r(E) = \sum_{p=L,R}\bold{\Sigma}_p^r(E)$, on the diagonal. All matrices labeled as $\check{\mathbf{O}}$ are representations of operators acting in the so-called Floquet-Sambe~\cite{Sambe1973} space, $\mathcal{H}_\mathrm{F} =  \mathcal{H}_\tau \otimes \mathcal{H}_e$, where $\mathcal{H}_e$ is the Hilbert space of electronic states spanned by localized orbitals $|i \rangle$ and $\mathcal{H}_\tau$ is the Hilbert space of periodic functions with period $\tau=2\pi/\omega_0$ spanned by orthonormal Fourier vectors $\langle t|n \rangle = \exp(i n \omega_0 t)$. Note that $\bold{1}$ is the unit matrix in $\mathcal{H}_e$ space and $\check{\bold{1}}$ is the unit matrix in $\mathcal{H}_\mathrm{F} =  \mathcal{H}_\tau \otimes \mathcal{H}_e$ space. 

The scattering matrix~\cite{Tserkovnyak2005} of the Landauer-B\"{u}ttiker approach to quantum transport, generalized to time-periodic multiterminal devices~\cite{Moskalets2011,Arrachea2006}, makes it possible to express the $N$th harmonic of spin current flowing into lead $p$  as~\cite{Moskalets2011} 
\begin{eqnarray}\label{eq:currentfloquet}
	I^{S_{\alpha}}_{p,N}  & = &  \frac{e}{\gamma} \hbar\omega_0 \sum_{p^{\prime}=L,R} \sum_{n=-\infty}^{\infty} n \mathrm{Tr}\bigg[ \hat{\sigma}_{\alpha} \mathbf{S}_{p p\prime}^{\dagger} \big(E_F+ n\hbar\omega_0,E_F\big) \nonumber \\
	&& \mathbf{S}_{pp\prime}\big(E_F+ (n+N)\hbar\omega_0,E_F\big) \bigg].
\end{eqnarray}
Equation~\eqref{eq:currentfloquet} is written in the limit of zero temperature and small frequency $\hbar \omega_0 \rightarrow 0$ which removes integrals~\cite{Moskalets2011} over energy and sets all quantities to depend only on the Fermi energy $E_F$. The charge current in lead $p$ is obtained from the same Eq.~\eqref{eq:currentfloquet} by replacement $\hat{\sigma}_{\alpha} \mapsto \hat{\sigma}_0$, where $\hat{\sigma}_0$ is the unit $2 \times 2$ matrix. Note that $I_{p,0}^{S_z}$ ($I_{p,0}$) is the DC component of spin (charge)  current, respectively.   The  Floquet scattering matrix~\cite{Moskalets2011,Arrachea2006} $\mathbf{S}_{p p^\prime}(E_F+n\hbar \omega_0,E_F)$ describes quantum transport of electrons at the Fermi energy from lead  $p'$ to $p$ while they absorb or emit  $n\hbar\omega_0$ ``photons.'' We compute $\mathbf{S}_{p p\prime}(E_F+n\hbar \omega_0,E_F)$ by using the generalization~\cite{Arrachea2006,FoaTorres2005,Kohler2005} of the Fisher-Lee formula,  originally derived for steady-state quantum transport~\cite{Fisher1981}, to  the case of periodically driven multiterminal devices
\begin{eqnarray}\label{eq:fisherlee}
	\lefteqn{\mathbf{S}_{p p\prime}(E_F+n\hbar\omega_0,E_F) = \delta_{p  p\prime}\delta_{E_F+n\hbar\omega_0,E_F}} \\
	 && -i\sqrt{\mathbf{\Gamma}_p(E_F+n \hbar \omega_0)} \mathbf{G}^r_{pp'}(E_F+n\hbar\omega_0,E_F)\sqrt{\mathbf{\Gamma}_{p\prime}(E_F)}, \nonumber 
\end{eqnarray}
Here $\mathbf{G}^r_{pp'}(E_F+n\hbar\omega_0,E_F)$ is the submatrix of Floquet retarded-GF  $\check{\bold{G}}^r(E)$ [Eq.~\eqref{eq:floquet_GF}], whose matrix elements connect sites along the edges of the central region that are also connected (via nonzero hopping) to the sites of the leads $p$ and $p'$; and   \mbox{${\bm \Gamma}_p(E) = i [{\bm \Sigma}_p^r(E) - {\bm \Sigma}_p^r(E)^\dagger]$} is the level broadening matrix~\cite{Stefanucci2013,Gaury2014,Popescu2016,Velev2004} of lead $p$. By truncating the infinite-dimensional space $\mathcal{H}_\tau$ to dimension $|n| \le n_\mathrm{max}$ we also convert infinite matrices $\check{\mathbf{O}}$ to finite-size ones suitable for numerical calculations, where convergence in $n$ is achieved by ensuring that each $N$th harmonic of pumped current in lead $p$  satisfies  $\bigg|\big(I_{p,N}(n_\mathrm{max}) - I_{p,N}(n_\mathrm{max}-1\big)/I_{p,N}(n_\mathrm{max}-1)\bigg|  < \delta$ with $\delta=10^{-2}$ chosen. This typically requires to use $n_\mathrm{max} \le 6$ for all harmonics $N$.

{\em Results and discussion.}---We warm up with calculations [flat dashed lines in Fig.~\ref{fig:fig2}] reproducing ``standard model'' results for spin pumping from FM~\cite{Tserkovnyak2005} or AFM~\cite{Cheng2014} in the absence of SOC, $\gamma_\mathrm{SO}=0$. As expected, our TDNEGF calculations reproduce time-independent $I_p^{S_z}$ and $I_p = 0$ (in the FM case) at sufficiently long times, after the transient response has died out.  The emergence of $I_p^{S_z}$ and $I_p \neq 0$ (in the AFM case) can be understood from the rotating frame picture of spin pumping~\cite{Chen2009,Hatori2007,Tatara2016,Tatara2019,Bajpai2020} where time-dependent setups in Fig.~\ref{fig:fig1} are mapped, by a unitary transformation into the frame which rotates with magnetization, onto four-terminal time-independent ones with effective bias voltage $\hbar \omega_0/e$~\cite{Chen2009,Tatara2016,Tatara2019} between the left or the right pairs of leads. In the rotating frame, $I_p^{S_z}$ is time-independent and it remains so upon transforming it into the lab frame, while $I_p=0$ in symmetric FM devices.  The {\em general requirement} for the appearance of nonzero DC component  of pumped charge current---that the left-right symmetry of a two-terminal device must be {\em broken}~\cite{Vavilov2001,FoaTorres2005,Bajpai2019}---helps to validate correctness of TDNEGF calculations. If this is done by breaking both inversion symmetry and time-reversal symmetry dynamically (such as by two spatially separated potentials oscillating out-of-phase~\cite{Switkes1999,Brouwer1998,Bajpai2019}) then DC component of charge current is nonzero and $\propto \omega_0$ at (sufficiently low) driving frequencies. If only one of those two  symmetries is broken, and this does not have to occur dynamically, then the DC component of pumped current  is $\propto \omega_0^2$ as the signature of nonadiabatic charge pumping~\cite{Vavilov2001,FoaTorres2005,Chen2009,Bajpai2019}. Since FM setup in Fig.~\ref{fig:fig1}(a) is left-right symmetric, DC component of pumped charge current in the presence of SOC is zero in Fig.~\ref{fig:fig2}(c). Conversely, AFM setup in  Fig.~\ref{fig:fig1}(b) has  broken (by configuration of magnetic moments) left-right symmetry, so DC component of its pumped charge current is nonzero in Fig.~\ref{fig:fig2}(d).

The rotating frame description becomes inapplicable when time-independent SOC is turned on in the lab frame because SOC becomes time-dependent in the rotating frame. Therefore, we switch to TDNEGF calculations in Fig.~\ref{fig:fig2} revealing that, as soon as the Rashba SOC is turned on, $I_p^{S_z}(t)$ oscillates harmonically and nonzero time-periodic $I_p(t)$ is established as well. Note that DC or perfectly harmonic currents are ensured in the long time limit by {\em continuous energy spectrum} of setups in Fig.~\ref{fig:fig1} brought by the attached NM leads and, thereby, dissipation effects generated by fermionic reservoirs.  

Figure~\ref{fig:fig3} shows that upon turning the Rashba SOC on, FFT power spectrum of pumped spin (note that spectra of $|I^{S_z}_R(\omega)|^2$ and $|I^{S_x}_R(\omega)|^2$ are nearly identical) and charge currents will also  contain high-harmonics at frequencies \mbox{$n=\omega/\omega_0$} for {\em both} even and odd integer $N$. Since only integer harmonics are present, in Fig.~\ref{fig:fig4} we switch to Floquet-NEGF formalism to show $|I_{R,N}|$ and $|I_{R,N}^{S_z}|$ vs. harmonic order $N$. This allows us to determine the cutoff harmonic order $N_\mathrm{max} \simeq 11$ with increasing SOC more precisely than when using TDNEGF calculations and FFT of their results (where apparently $N_\mathrm{max} \simeq 4$ in Fig.~\ref{fig:fig3}) where numerical artifacts are easily introduced by the choice of time step and FFT window.

The microscopic origin of the transition from time-independent $I_p^{S_z}$ for $\gamma_\mathrm{SO}=0$ to time-dependent $I_p^{S_z}(t)$ for $\gamma_\mathrm{SO}\neq 0$ can be understood from animation, provided as the  Supplemental Material (SM)~\cite{sm}, of nonequilibrium spin density \mbox{$\langle \hat{\mathbf{s}}_1 \rangle^\mathrm{neq}(t) = \mathrm{Tr}\, \big[|1 \rangle \langle 1| \otimes \hat{\bm \sigma} {\bm \rho}^{\rm neq}(t) \big]$} at site 1 in the FM case. While \mbox{$\langle \hat{s}_i^z \rangle^\mathrm{neq}$} is time-independent for \mbox{$\gamma_\mathrm{SO}=0$}~\cite{Tserkovnyak2005,Costache2008}, it becomes harmonically time-dependent with frequency $\omega_0$ and its integer multiples, so that \mbox{$\langle \hat{\mathbf{s}}_i \rangle^\mathrm{neq}(t)$} nutates as it flows out of FM to comprise~\cite{Costache2008} pumped spin current and give rise to  $I_p^{S_z}(t)$. 

The predicted high-harmonic spectra [Fig.~\ref{fig:fig4}] for spin and charge currents pumped by precessing magnetization bear resemblance to a field where high-harmonic generation in pumped charge current has been intensely pursued in recent years---solids driven out of equilibrium by laser light of frequency $\omega_0$~\cite{Ghimire2019}. For example, inversion symmetric bulk semiconductors driven by strong mid-infrared laser light, whose $\hbar \omega_0$ is much smaller than the band gap, can exhibit nonlinear effect generating new radiation at odd multiplies of $\omega_0$~\cite{Ghimire2019}. Furthermore, in two-dimensional (2D) systems breaking inversion symmetry, such as monolayer of MoS$_2$~\cite{Ghimire2019} or surface states of topological materials~\cite{Bai2021},  additional even-order harmonics or non-integer harmonics~\cite{Bai2021} can emerge. This has inspired recent theoretical studies on possible high-harmonic generation in spin currents pumped by laser light irradiating magnetic insulators~\cite{Ikeda2019,Takayoshi2019}; as well as in SO-split 3D materials~\cite{Tancogne-Dejean2022} and 2D electron gases~\cite{Lysne2020}. However, these schemes rely on highly nonlinear effects in strong light-matter coupling, and in the case of magnetic insulators they assume coupling of magnetic field of laser light directly to localized magnetic moments. Since such coupling is $1/c$ smaller than light-charge coupling, they would require intense THz laser pulses beyond presently available technologies.

On the other hand, setups in Fig.~\ref{fig:fig1} are routinely made in spintronics using widely available microwave sources and, in contrast to optical pumping, with {\em low input power} $\sim$ mW~\cite{Fan2010} and possibility for scalability. For example, first-principles Floquet-NEGF  analysis~\cite{Dolui2022} of the very recent experiments~\cite{Vaidya2020} on spin pumping from AF insulator MnF$_2$ in contact with heavy metal Pt has revealed that MnF$_2$ layer can be significantly modified by SO-proximity effect~\cite{Zutic2019,Marmolejo-Tejada2017,Dolui2020b,Dolui2020a,Dolui2020,Dolui2022} within such heterostructure due to SOC at interfaces or from the bulk of Pt layer. This means that re-examining such experiments, where all key ingredients for our predictions are already present, could reveal high-harmonics in electromagnetic radiation produced by pumped time-dependent charge current [Figs.~\ref{fig:fig2}(d) and ~\ref{fig:fig3}(d)] or spin current [Figs.~\ref{fig:fig2}(b) and ~\ref{fig:fig3}(b)] converted to charge current~\cite{Wei2014} by Pt layer. 

Furthermore, one could search for F or AF materials with strong intrinsic SOC~\cite{Ciccarelli2015}. In this respect, 2D magnetic materials~\cite{Gibertini2019} are particularly promising choice since their magnetic ordering at finite temperature crucially relies on magnetic anisotropy originating from strong SOC~\cite{Olsen2019,Vanherck2020}. In addition, 2D magnetic materials can be easily SO-proximitized~\cite{Zutic2019} by transition metal dichalcogenides to further tune their properties~\cite{Dolui2020}. Thus, taking into account that 2D magnetic materials are typically of honeycomb or hexagonal lattice type~\cite{Gibertini2019,Olsen2019,Vanherck2020}, we examine a 2D setup in Fig.~\ref{fig:fig1}(c) where the honeycomb lattice hosting both precessing magnetic moments and the Rashba SOC [Eq.~\eqref{eq:hamil2d}] is attached to semi-infinite GNR leads. The  pumped spin [Fig.~\ref{fig:fig5}(a)] and charge [Fig.~\ref{fig:fig5}(b)] currents in such model of 2D magnet at F resonance exhibit high harmonics with larger cutoff  $N_\mathrm{max} \simeq 25$ 
than in the case of 1D setups studied in Fig.~\ref{fig:fig4}. 

{\em Conclusions}.---Using TDNEGF calculations~\cite{Gaury2014,Popescu2016}, applicable to arbitrary time-dependent quantum transport of spin and charge in multi-terminal devices~\cite{Petrovic2018,Bajpai2019,Petrovic2021}, we find that periodically driven by microwaves  magnetic moments of FM or AFM materials will 
pump spin and charge currents oscillating at both frequency $\omega_0$ of the driving field and its integer high harmonics $N\omega_0$. This is in contrast to two-decades of intense studies~\cite{Tserkovnyak2005} of current pumping in spintronics by dynamical magnetization where only spin current $I_p^{S_z}$ is found with one of its three components being DC and the other two oscillating with frequency $\omega_0$. We additionally employ Floquet-NEGF formalism, combining the Floquet-scattering-matrix~\cite{Moskalets2011} with time-independent NEGF calculations~\cite{Mahfouzi2012}, which allows us to precisely estimate [Figs.~\ref{fig:fig4} and ~\ref{fig:fig5}] cutoff harmonic $N_\mathrm{max}$, as well as validate TDNEGF calculations (that are 
initially deployed to unearth any processes beyond Floquet-based formalism, such as non-integer harmonics). Our prescription for experimental realization of 
this effect is based [Figs.~\ref{fig:fig1}(c) and ~\ref{fig:fig5}] on 2D FM or AFM materials~\cite{Ghimire2019,Olsen2019,Vanherck2020} which have intrinsically strong SOC that can be easily further tailored by proximity effect~\cite{Zutic2019} to nonmagnetic 2D materials~\cite{Dolui2020} within van der Waals heterostructures. Finally, using AF materials driven into resonance by $\omega_0$ in sub-THz range~\cite{Vaidya2020,Li2020} means that $I_p(t)$ they pump would generate output THz radiation at multiples of the driving frequency $\omega_0$, thereby opening new avenues for THz spintronics where such output radiation is presently generated by more complex F~\cite{Seifert2016} or AF  heterostructures~\cite{Qiu2021,Suresh2022} driven by femtosecond laser pulses.

{\em Note added.}---This study was initially motivated by an attempt to confirm TDNEGF calculations of Ref.~\cite{Ly2022}. However,  all results of Ref.~\cite{Ly2022} for finite SOC conspicuously violate two theorems of time-dependent quantum transport, as discussed in detail in~\cite{Nikolic2022}, and 
even if TDNEGF calculations of Ref.~\cite{Ly2022} had been correct, it would have been virtually impossible [Fig.~\ref{fig:fig3} vs. Fig.~\ref{fig:fig4}] to determine precisely $N_\mathrm{max}$ without involving Floquet-NEGF [Eqs.~\eqref{eq:doubleft}--\eqref{eq:fisherlee}] calculations as well.  

\begin{acknowledgments}
This work was supported by the US National Science Foundation (NSF) through the University of Delaware Materials Research Science and Engineering Center DMR-2011824.
\end{acknowledgments}


\end{document}